\documentstyle[twocolumn,aps,prl,floats,epsf,psfig]{revtex}
\begin{document}

\draft

%* comment this line out for ``long'' draft format
\wideabs{

\title{DILEPTON PRODUCTION AT FERMILAB AND RHIC}

\author{
J. C.~Peng, P. L.~McGaughey, and J. M.~Moss}
\vskip 0.5cm
\address{
Physics Division, Los Alamos National Laboratory, Los Alamos, NM 87545}

\maketitle
\begin{abstract}
Some recent results from several fixed-target dimuon production experiments 
at Fermilab are presented. In particular, we discuss the use of Drell-Yan
data to determine the flavor structure of the nucleon sea, as well as
to deduce the energy-loss of partons traversing nuclear medium. 
Future dilepton experiments at RHIC could shed more light on
the flavor asymmetry and possible charge-symmetry-violation of the 
nucleon sea. Clear evidence for scaling violation in the Drell-Yan
process could also be revealed at RHIC.
\end{abstract} 
}

\noindent{\bf 1. Introduction}\\

\noindent The first dilepton production experiment was performed at the AGS
almost 30 years ago~\cite{leon}. Over the years, 
hadron-induced dilepton production
experiments have led to the discoveries of various vector bosons ($J/\Psi$,
$\Upsilon$, and Z$^0$). They also provided important and often unique
informations on parton distributions in nucleons, nuclei, and mesons.
 
A series of fixed-target dimuon production experiments (E772, E789, E866)
have been carried out at Fermilab in the last 10 years. Some highlights
from these experiments are presented here. In addition, the prospect for
performing several dilepton production experiments at RHIC is discussed.\\

\noindent{\bf 2. Scaling Violation in Drell-Yan Process}\\

\noindent In the ``Naive" Drell-Yan (DY) model,
the differential cross section, $m^3d^2\sigma/dx_Fdm$, has an expression showing
its scaling property as follows:
\begin{eqnarray}
\lefteqn{M^3 {d^2\sigma\over dM dx_F} = {8\pi\alpha^2\over 9}
{x_1 x_2\over x_1 + x_2} \times }\hspace{0.1in} \nonumber \\
 & \sum_a e_a^2[q_a(x_1)
\bar q_a(x_2)+\bar q_a(x_1)q_a(x_2)]. \label{eq:dy}
\end{eqnarray}
The right-hand side of Eq. (1) is only a function of $x_1, x_2$ and is 
independent of the beam energies. This scaling property no longer holds
when QCD corrections to the DY are taken into account.

While logarithmic scaling violation is well established in 
Deep-Inelastic Scattering (DIS) experiments,
it is not well confirmed in DY experiments at all. 
As an example, Figure 1 compares the NA3 data~\cite{na3} at 
400 GeV with the E605~\cite{e605} and E772~\cite{e772a} data at 800 GeV. 
The solid curve in Figure 1 corresponds to NLO calculation 
for 800 GeV $p+d$ ($\sqrt s$ = 38.9 GeV) and it describes
the NA3/E605/E772 data well. No evidence for
scaling violation is seen. As discussed in a recent review~\cite{plm}, 
there are mainly two reasons for this. First, unlike the DIS, 
the DY cross section is a convolution of two structure functions. Scaling 
violation implies that the structure functions rise for $x \leq 0.1$ and 
drop for $x \geq 0.1$ as $Q^2$ increases. For proton-induced
DY, one often involves a beam quark with $x_1 > 0.1$ and a target antiquark
with $x_2 < 0.1$. Hence the effects of scaling violation are partially 
cancelled. Second, unlike the DIS, the DY experiment can only probe 
relatively large $Q^2$, namely, $Q^2 > 16$ GeV$^2$ for a mass cut of 4 GeV. 
This makes it more difficult to observe the logarithmic variation of the 
structure functions in DY experiments.

Possible indications of scaling violation in DY process have been reported
in two pion-induced experiments, E326~\cite{e326} at 
Fermilab and NA10~\cite{na10} at CERN.
E326 collaboration compared their 225 GeV $\pi^- + W$ DY cross 
sections against calculations with and without scaling violation.
They observed better agreement when scaling violation is included.
This analysis is subject to the uncertainties associated with 
the pion structure functions, as well as the nuclear effects of the $W$ target.
The NA10 collaboration measured $\pi^- + W$ DY cross sections at three
beam energies, namely, 140, 194, and 286 GeV. By checking the ratios of the 
cross sections at three different energies, NA10 largely avoids the 
uncertainty of the pion structure functions. However, the relatively small 
span in $\sqrt s$, together with the complication of nuclear effects, make 
the NA10 result less than conclusive.

RHIC provides an interesting opportunity for unambiguously establishing 
scaling violation in the DY process. 
Figure 1 shows the predictions for $p+d$ at $\sqrt s$ = 500 GeV. 
The scaling-violation accounts for a factor of two
drop in the DY cross sections when $\sqrt s$ is increased from 38.9 GeV
to 500 GeV. It appears quite feasible to establish scaling violation in
DY with future dilepton production experiments at RHIC.\\

\noindent{\bf 3. Flavor-Asymmetry and Charge-Symmetry-Violation
of the Nucleon Sea}\\

\noindent The DY process complements DIS as a tool to probe parton distributions
in nucleons and nuclei. This is well illustrated in the recent progress
in the study of flavor-asymmetry in the nucleon sea.

Until recently, it had been assumed that the distributions of $\bar u$ 
and $\bar d$ quarks in the proton were identical. 
While the equality of $\bar u$ and 
$\bar d$ is not required by any known symmetry, it is a plausible
assumption for sea quarks generated by gluon splitting. As the masses of the 
up and down quarks are small compared to the confinement scale, nearly equal 
number of up and down sea quarks should result.

\begin{figure}
\begin{center}
\psfig{figure=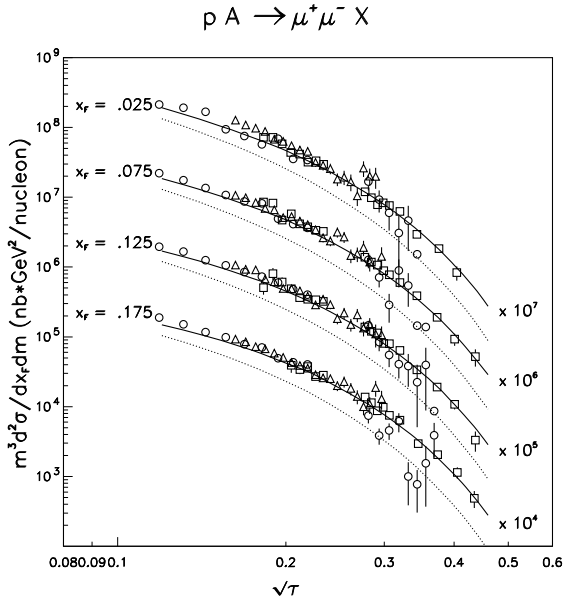,height=3.2in}
\end{center}
\caption{Comparison of DY cross section data with NLO
calculations using MRST [15] structure 
functions. Note that $\tau = x_1 x_2$.
The E772 [4], E605 [3], and
NA3 [2] data points are shown as circles, 
squares, and triangles, respectively.
The solid curve corresponds to fixed-target 
p+d collision at 800 GeV, while the
dotted curve is for p+d collision at $\sqrt s$ = 500 GeV.}
\label{fig:fig1.eps}
\end{figure}

The assumption of $\bar u(x) = \bar d(x)$ can be tested by measurements
of the Gottfried integral~\cite{gott}, defined as
\begin{eqnarray}
\lefteqn{I_G = \int_0^1 \left[F^p_2 (x,Q^2) - F^n_2 
(x,Q^2)\right]/x~ dx =}\hspace{0.7in} \nonumber \\
 & {1\over 3}+{2\over 3}\int_0^1 \left[\bar u_p(x)-\bar d_p(x)\right]dx,
\label{eq:3.1}
\end{eqnarray}
where $F^p_2$ and $F^n_2$ are the proton and neutron structure
functions measured in DIS experiments. The second step in 
Eq.~\ref{eq:3.1} follows from the assumption of
charge symmetry.
Under the assumption of a symmetric sea, $\bar u_p$ = $\bar d_p$,
the Gottfried Sum Rule (GSR)~\cite{gott}, $I_G
= 1/3$, is obtained. 

The most accurate test of the GSR was reported by the New Muon 
Collaboration (NMC)~\cite{nmc}, which measured $F^p_2$ and $F^n_2$ over the 
region $0.004 \le x \le 0.8$. They determined the Gottfried integral to be 
$ 0.235\pm 0.026$, significantly below 1/3. This result implies that
the integral of $\bar d -\bar u$ is nonzero. However, the $x$-dependence
of $\bar d -\bar u$ remained unspecified.

The proton-induced DY process provides an
independent means to probe the flavor asymmetry of the nucleon sea~\cite{es}.
An important advantage of the Drell-Yan process is that the $x$-dependence of 
$\bar d / \bar u$ can be determined.
The NA51~\cite{na51} and the E866~\cite{e866} 
collaborations have compared the proton-induced 
DY dimuon yields from hydrogen and deuterium targets, and they deduced
the ratios of $\bar d/\bar u$ as shown in Figure 2.
For $x < 0.15$, $\bar d/\bar u$ increases linearly with $x$ and is in
good agreement with the CTEQ4M~\cite{cteq} and MRS(R2)~\cite{mrs} parameterizations. 
However, a distinct feature of the data, not seen in either 
parameterization, is the
rapid decrease towards unity of $\bar{d}/\bar{u}$ beyond
$x_{2}=0.2$. 
The E866 data clearly affect the current parameterization of the
nucleon sea. The most recent 
structure functions of Martin et al.~\cite{mrst}(MRST)
included the E866 data in its global fit and 
its parametrization of $\bar d / \bar u$ is shown in Figure 2.

\begin{figure}
\begin{center}
\psfig{figure=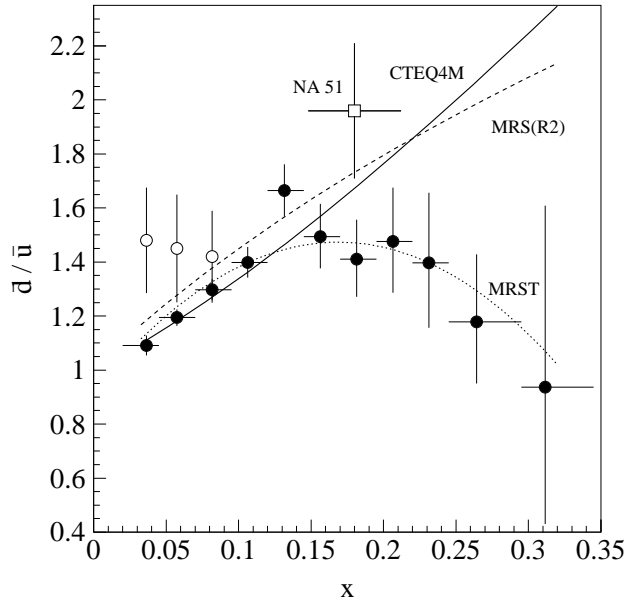,height=3.2in}
%\mbox{\epsffile{jcp_fig2.eps}}
\end{center}
\vskip -0.3cm
  \caption{The ratio of $\bar{d}/\bar{u}$ in the proton as a function
  of $x$ extracted from the Fermilab E866 cross section ratio. The
  curves are from various parton distributions.  The error bars
  indicate statistical errors only.  An additional systematic
  uncertainty of $\pm0.032$ is not shown.  Also shown is the result
  from NA51, plotted as an open box. The open circles correspond
  to $\bar d/\bar u$ values extracted with the assumption of the CSV
  effect reported in ref. [10].}
\label{fig:fig2}
\end{figure}

Many papers have considered virtual mesons as the origin for the
observed  $\bar d/\bar u$ asymmetry (see recent review of 
Kumano~\cite{kumano0}). Here the $\pi^+(\bar d u)$ cloud, dominant in the 
process, $p\rightarrow\pi^+ n$, leads to an excess of $\bar d$ sea.
Comparison of the E866 data with various theoretical models has
also been made~\cite{peng}.

It should be emphasized that the extraction of $\bar d / \bar u$ values from the
NA51 and E866 DY experiments required the assumption of charge symmetry, 
namely, $\bar d_n = \bar u_p, \bar u_n = \bar d_p$, etc.
Evidence for a surprisingly large charge-symmetry-violation (CSV) 
effect was recently reported by
Boros et al.~\cite{boros1,boros2} 
based on an analysis of $F_2$ structure functions
determined from muon and neutrino DIS experiments. A large asymmetry,
$\bar d_n(x) \approx 1.25 \bar u_p(x)$ for $0.008 < x < 0.1$, is apparently
needed to bring the muon and neutrino DIS data into agreement.
How would this finding, if confirmed by further studies, affect the 
E866 analysis of the flavor asymmetry? First, CSV alone could
not account for the E866 data. In fact, an even larger
amount of flavor asymmetry is required to compensate the possible CSV 
effect~\cite{boros2}. This is illustrated in Figure 2, where the open circles
correspond to the $\bar d/ \bar u$ values one would have obtained if the
CSV effect reported by Boros et al. is assumed. 
Second, there has been no indication of CSV for $x > 0.1$. Thus the
large $\bar d / \bar u$ asymmetry from E866 for $x > 0.1$
is not affected. 

\begin{figure}
\begin{center}
\psfig{figure=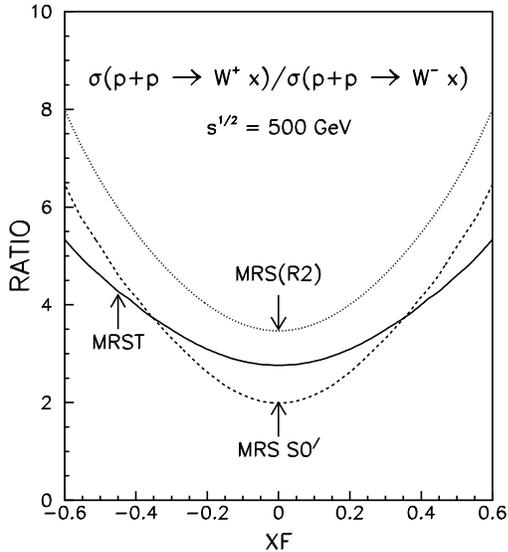,height=3.2in}
\end{center}
\vskip -0.3cm
  \caption{ Predictions of 
  $\sigma (p+p \to W^+ x) / \sigma (p+p \to W^- x)$ as a function of $x_F$
  at $\sqrt s$ = 500 GeV.
  The dashed curve corresponds to the $\bar
  d/\bar u$ symmetric MRS S0$^\prime$ structure 
  functions, while the solid and dotted curves
  are for the $\bar d/\bar u$ asymmetric structure function MRST and MRS(R2),
  respectively.}
\label{fig:fig3}
\end{figure}

To disentangle the $\bar d / \bar u$ asymmetry from the possible 
CSV effect, one could consider $W$ boson production, a generalized
DY process, in $p + p$ collision at RHIC.
An interesting quantity to be measured is the ratio of the 
$p + p \to W^+ + x$ and $p + p \to W^- + x$ cross sections~\cite{peng1}. 
It can be shown that this
ratio is very sensitive to $\bar d / \bar u$. An important feature of
the $W$ production asymmetry in $p + p$ collision is that it is completely free 
from the assumption of charge symmetry. Figure 3 shows the 
predictions for $p + p$ collision at $\sqrt s =
500~$GeV. The dashed curve corresponds to the $\bar
d/\bar u$ symmetric MRS S0$^\prime$~\cite{mrss0} structure 
functions, while the solid and dotted curves
are for the $\bar d/\bar u$ asymmetric structure function MRST and MRS(R2),
respectively.
Figure 3 clearly shows that $W$ asymmetry measurements at RHIC could 
provide an independent determination of $\bar d / \bar u$.\\

\noindent{\bf 4. Nuclear Medium Effects of Dilepton Production}\\

\noindent From a high-statistics measurement of dilepton production in 800 GeV
proton-nucleus interaction, the target-mass dependence of DY, $J/\Psi$,
$\Psi^\prime$, and $\Upsilon$ productions have 
been determined in E772~\cite{e772b,e772c,e772d}.
As shown in Figure 4, different nuclear dependences are observed for
different dilepton processes. While the DY process shows almost no nuclear
dependence, pronounced nuclear effects are seen for the production of
heavy quarkonium states. E772 found that $J/\Psi$ and $\Psi^\prime$
have similar nuclear dependence. The nuclear dependences for $\Upsilon$,
$\Upsilon^\prime$ and $\Upsilon^{\prime \prime}$ are less than that observed
for the $J/\Psi$ and $\Psi^\prime$. Within statistics, the various 
$\Upsilon$ resonances also have very similar nuclear dependences.

\begin{figure}
\begin{center}
\psfig{figure=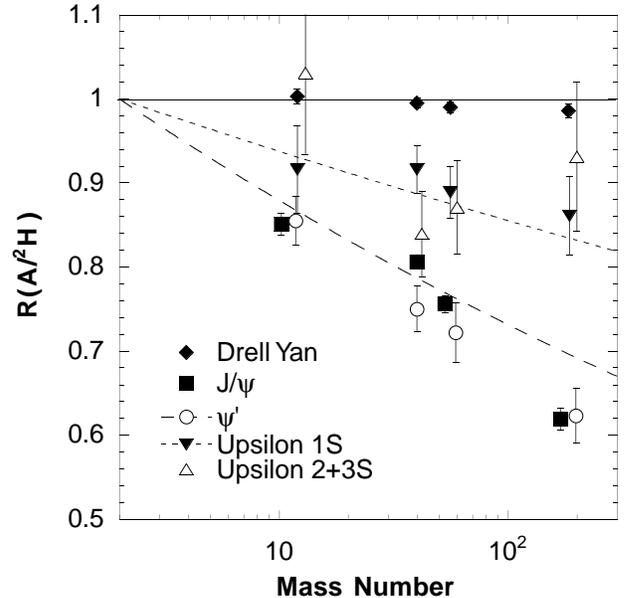,height=3.2in}
%\mbox{\epsffile{rhicfig4.eps}}
\end{center}
\vskip -0.3cm
\caption{ Ratios of heavy-nucleus to deuterium
integrated yields per nucleon for 800 GeV proton production of dimuons from the 
Drell-Yan process and from decays of the $J/\psi$, $\psi '$, $\Upsilon 
(1S)$, and $\Upsilon (2S+3S)$ states [5]. The short 
dash and long dash 
curves represent the approximate nuclear dependences for the $b\bar b$ 
and $c\bar c$ states, $A^{0.96}$ and $A^{0.92}$, respectively.}
\label{fig:fig4}
\end{figure}

\begin{figure}
\begin{center}
\psfig{figure=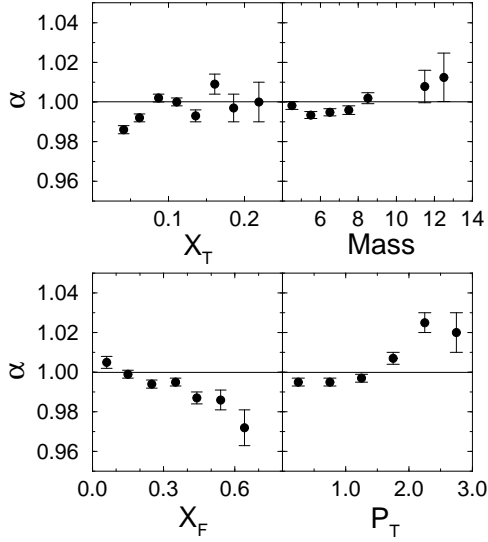,height=3.8in}
%\mbox{\epsffile{rhicfig1.eps}}
\end{center}
\vskip -1.0cm
  \caption{Nuclear dependence coefficient $\alpha$ for 800 GeV p+A
  Drell-Yan process versus various kinematic variables [22].}
\label{fig:fig5}
\end{figure}

\begin{figure}
\begin{center}
\psfig{figure=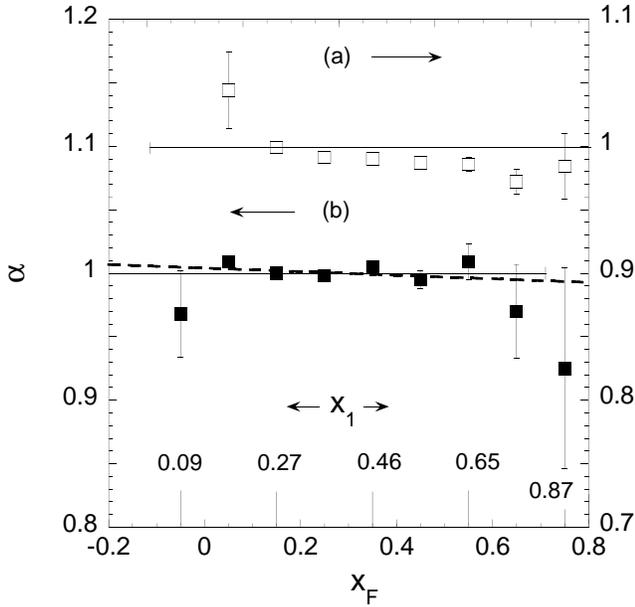,height=3.2in}
%\mbox{\epsffile{rhicfig6.eps}}
\end{center}
\vskip -0.3cm
\caption{  Nuclear dependence coefficient $\alpha$ for the Drell-Yan 
process [22] versus
$x_F$ for (a) $x_2\leq 0.075$, right scale, and  (b) $x_2\geq 0.075$, left 
scale. The thin solid lines show $\alpha =1$. The dashed line is a 
linear least-squares fit to the lower points. Also shown is the 
mean value of $x_1$ for (b).}
\label{fig:fig6}
\end{figure}

\begin{figure}
\begin{center}
\psfig{figure=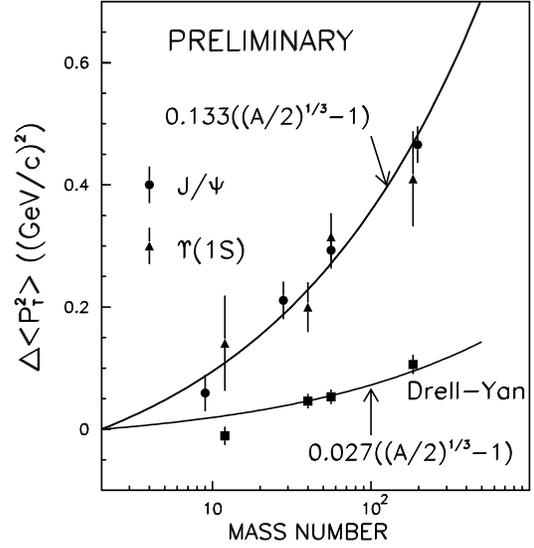,height=3.2in}
%\mbox{\epsffile{rhicfig7.eps}}
\end{center}
\vskip -0.3cm
  \caption{The change of mean $p_t^2$ for nuclear target, 
  $\Delta \langle p_t^2 \rangle = \langle p_t^2 \rangle (A) -
  \langle p_t^2 \rangle (D)$, for 800 GeV p+A Drell-Yan process and
  $J/\Psi$ and $\Upsilon(1S)$ productions. The solid curves are best
  fits to the A-dependence of $\Delta \langle p_t^2 \rangle$. The
  $J/\Psi$ and $\Upsilon (1S)$ productions have identical 
  curves for the A-dependence
  fits.}
\label{fig:fig7}
\end{figure}

Although the integrated DY yields in E772 show little nuclear dependence,
it is instructive to examine the DY nuclear dependences on
various kinematic variables. Using the simple $A^\alpha$ expression 
to fit the DY nuclear dependence, the values of $\alpha$ are shown in Figure 5 
as a function of $x_T (x_2)$, $M$, $x_F$, and $p_t$.
Several features are observed:

\begin{enumerate}

\item A suppression of the DY yields from heavy nuclear targets is seen
at small $x_2$. This is consistent with the shadowing effect observed
in DIS. In fact, E772 provides the only experimental evidence for
shadowing in hadronic reactions. The reach of small $x_2$
in E772 is limited by the mass cut ($M \geq 4$ GeV) and by the
relatively small center-of-mass energy (recall that $x_1 x_2 = M^2/s$). p-A
collisions at RHIC clearly offer the exciting opportunity to extend the
study of shadowing to much smaller $x$.

\item $\alpha (x_F)$ shows an interesting trend, namely, it decreases
as $x_F$ increases. It is tempting to attribute this behavior to 
initial-state energy-loss effect. However, there is a strong correlation
between $x_F$ and $x_2$ ($x_F = x_1 -x_2$), and it is essential to
separate the $x_F$ energy-loss effect from the $x_2$ shadowing effect. 
Figure 6 shows $\alpha$ versus $x_F$ for two bins of $x_2$, one in the
shadowing region ($x_2 \leq 0.075$) and one outside of it ($x_2 \geq 0.075$).
There is no discernible $x_F$ dependence for $\alpha$ once one stays outside
of the shadowing region. Therefore, the 
apparent suppression at large $x_F$ in Figure 5
reflects the shadowing effect at small $x_2$ rather than the energy-loss effect.

\item $\alpha (p_t)$ shows an enhancement at large $p_t$. This is reminiscent
of the Cronin Effect~\cite{cronin} where the broadening in $p_t$ distribution
is attributed to multiple parton-nucleon scatterings. It is instructive to 
compare the $p_t$ broadening for DY process and quarkonium production.
Figure 7 shows $\Delta\langle p_t^2\rangle$, 
the difference of mean $p_t^2$ between
p-A and p-D interactions, as a function of A for DY, J/$\Psi$, 
and $\Upsilon(1S)$
productions at 800 GeV. The DY and $\Upsilon$ data 
are from E772~\cite{e772e}, while the
$J/\Psi$ results are from E789~\cite{e789}, E771~\cite{e771}, and
preliminary E866 analysis~\cite{leitch}. 
More details on this analysis will be 
presented elsewhere~\cite{e772e}. Figure 7 shows 
that $\langle p_t^2\rangle$ is well described by
the simple expression $a + b A^{1/3}$. It also shows that the
$p_t$ broadening for $J/\Psi$ is very similar to $\Upsilon$,
but significantly larger (by a factor of 5) than the DY. A factor of
9/4 could be attributed to the color factor of the initial gluon in
the quarkonium production versus the quark in the DY process. 
The remaining difference
could come from the final-state multiple scattering effect which is absent
in the DY process.

\end{enumerate}
 
Baier et al.~\cite{baier1} have 
recently derived a relationship between the partonic 
energy-loss due to gluon bremsstrahlung and the mean $p_t^2$ broadening
accumulated via multiple parton-nucleon scattering:
\begin{eqnarray}
-dE/dz = {3\over 4}~ \alpha_s~ \Delta\langle p_t^2\rangle. \label{eq:dedx}
\end{eqnarray}
This non-intuitive result states that the total energy loss
is proportional to square of the path length traversed by the incident partons.
From Figure 7 and Eq.~\ref{eq:dedx}, we deduce that the mean total energy loss,
$\Delta E$, for the p+W
DY process is $\approx 0.6$ GeV. Such an energy-loss is too small to
cause any discernible effect in the $x_F$ (or $x_1$) nuclear dependence.
As shown in Figure 6, the dashed curve corresponds 
to $\Delta E = 2.0 \pm 1.7$ GeV (for p+W), and the E772 data are consistent with
Eq.~\ref{eq:dedx}. A much more sensitive test for Eq.~\ref{eq:dedx} could 
be done at RHIC, where the energy-loss effect is expected to be much enhanced
in A-A collision~\cite{baier2}.

\end{document}